# *Analysis of The Anomalous Orbital-Energy Changes Observed in Spacecraft Flybys of Earth*

*by*

*Roger Ellman*

Abstract

In March 2008 anomalous behavior in spacecraft flybys of Earth was reported in Physical Review Letters, Volume 100, Issue 9, March 7, 2008, in an article entitled "Anomalous Orbital-Energy Changes Observed during Spacecraft Flybys of Earth"[1].

The data indicate unaccounted for changes in spacecraft speed, both increases and decreases, for six different spacecraft involved in Earth flybys from December 8, 1990 to August 2, 2005. The article states that, "All … potential sources of systematic error …. [have been] modeled. None can account for the observed anomalies…. Like the Pioneer anomaly … the Earth flybys anomaly is a real effect …. Its source is unknown."

In the present article it is shown that the Earth flybys anomaly would be caused by a very small acceleration [in addition to that of natural gravitation], centrally directed and independent of distance, the same effect as that which the Pioneer anomaly exhibits. How that effect operates to produce the observed results is analyzed. A cause of the centrally directed accelerations is presented.

Roger Ellman,  The-Origin Foundation, Inc.
               320 Gemma Circle, Santa Rosa, CA 95404, USA
               RogerEllman@The-Origin.org
               707-537-0257
               http://www.The-Origin.org

# *Analysis of The Anomalous Orbital-Energy Changes Observed in Spacecraft Flybys of Earth*

*by*
*Roger Ellman*

## *The Problem*

In March 2008 anomalous behavior in spacecraft flybys of Earth was reported in Physical Review Letters, Volume 100, Issue 9, March 7, 2008, in an article entitled "Anomalous Orbital-Energy Changes Observed during Spacecraft Flybys of Earth"[1].

The data indicate unaccounted for changes in spacecraft speed, both increases and decreases, for six different spacecraft involved in Earth flybys from December 8, 1990 to August 2, 2005. The article states that, "All … potential sources of systematic error …. [have been] modeled. None can account for the observed anomalies…. "Like the Pioneer anomaly … the Earth flybys anomaly is a real effect …. Its source is unknown."

## *Analysis of the Flyby Anomaly*

The phenomenon that would account for the highly varied occurrences of the flyby anomaly is a small acceleration [in addition to that of natural gravitation], centrally directed and independent of distance; that is a modest and otherwise unknown acceleration directed toward the core center of the Earth, the principle body involved in the mechanics of the flyby.

Such an acceleration shares characteristics with the Pioneer anomaly, which appears as a small, centrally directed, distance independent acceleration, that is a modest and otherwise unknown acceleration directed toward the Sun, the principle body involved in the mechanics of the Pioneer spacecrafts' flight paths. Thus the two effects, the Flyby Anomaly and the Pioneer Anomaly would appear to share the same cause based on the analysis below of the Flybys Anomaly. The Pioneer anomaly is thoroughly analyzed and explained in *A Comprehensive Resolution of the Pioneer 10 and 11 "Anomalous Acceleration" Problem Presented in the Comprehensive Report "Study of the Anomalous Acceleration of Pioneer 10 and 11" and the Relationship of that Issue to "Dark Matter", "Dark Energy", and the Cosmological Model*[2].

In the referenced Pioneer Anomaly article the centrally directed, distance independent, Pioneer acceleration is shown to be an effect that concomitantly involves centrally directed, distance independent, acceleration in all cosmic structures. Thus its occurrence in the Earth flyby cases is to be expected.

To observe the relation to the Flybys Anomaly of an otherwise unknown or un-detected anomalous, centrally directed, distance independent acceleration the first step is to consider a simple spacecraft pass of Earth where the pass is all at zero latitude as shown in Figure 1, on the following page. In the vectors analysis part of the figures *A* is the full anomalous acceleration, *C* is its component parallel to the direction of motion of the satellite, and *θ* is the angle between the direction of action of those two.

When the spacecraft is at a great distance out from Earth the spacecraft's motion is close to being directed toward the center of the Earth but not exactly so. A centrally directed acceleration there analyzed into components parallel and perpendicular to the spacecraft's motion would show most of the centrally directed acceleration acting to increase the spacecraft's speed.

a. <u>Polar View - Flyby</u>

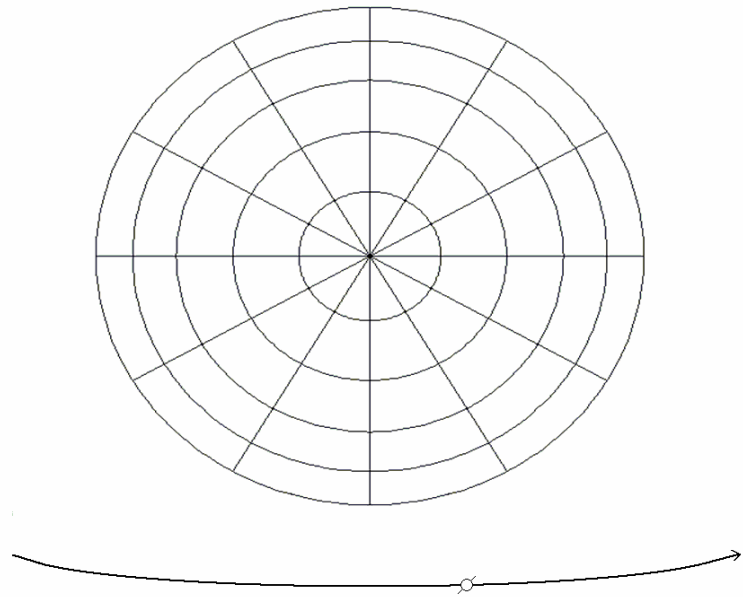

b. <u>Polar View - Anomalous Acceleration Vectors</u>

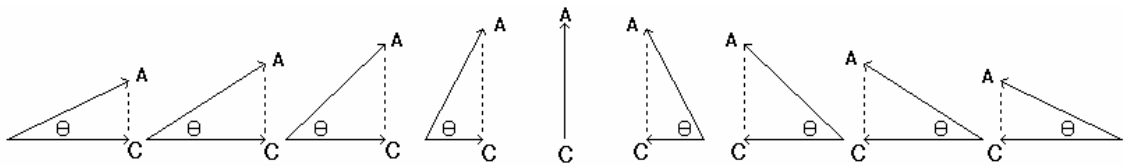

Here the acceleration phase and the deceleration phase are equal and offset each other.

b. <u>Equatorial View</u>

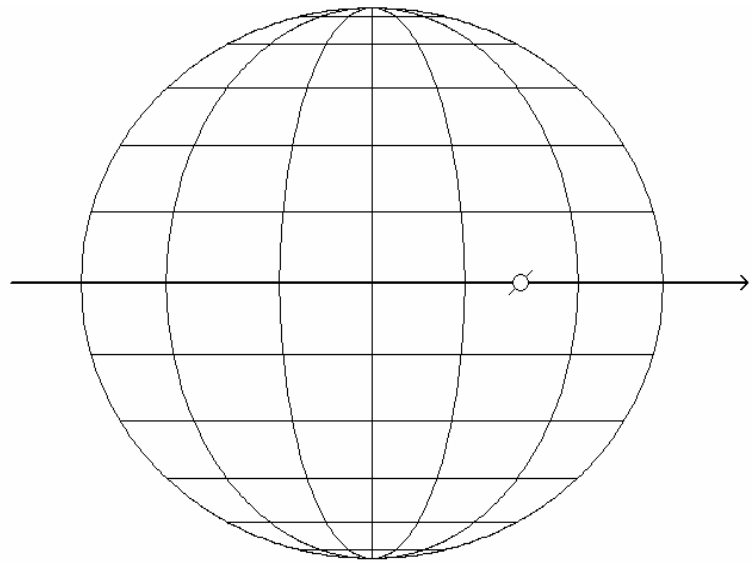

*Figure 1*
*A Zero Latitude Pass*

As the spacecraft travels nearer to Earth that component parallel to its motion decreases, becoming zero at the closest approach to Earth. From that point on the parallel component acts in the opposite direction on the spacecraft, that is its effect is to decelerate the spacecraft not accelerate it. Ultimately the anomalous acceleration and anomalous deceleration experienced by the spacecraft become equal and cancel each other out leaving as the only flyby effect the gravitational boost, due to another effect, that is the overall purpose of the flyby.

Of the full centrally directed acceleration, $A$, the component, $C$, parallel to the path of the flyby in this case is

*(1)*   `C = A·Cos[θ]`

which is apparent if the flyby path is a straight line. However, the actual flyby path is somewhat curved by the Earth's gravitation. But, the anomalous acceleration is always centrally directed toward the core of the Earth so that $C$ is nevertheless as stated.]

Equation *(1)* is valid when the flyby pass is solely at zero latitude. However, if other than zero the latitude of the flyby pass has a significant effect on the magnitude of $C$, the component of the overall centrally directed acceleration parallel to the spacecraft flight path. As latitude increases the magnitude of $C$, decreases. That is most easily visualized by imagining the flyby over the geographic north pole at 90° north latitude. There the centrally directed acceleration toward the center of the Earth has no component parallel to the flight path.

Therefore, for flyby paths at other than zero latitude the effective value of $A$ is $A(\lambda)$ a function of latitude, $\lambda$, as equation *(2)*

*(2)*   `A = A(λ) = A·Cos[λ]`

so that equation *(1)* then becomes equation *(3)* the full expression for the extent to which the centrally directed anomalous acceleration actually accelerates or decelerates the spacecraft.

*(3)*   `C = A·Cos[λ]·Cos[θ]`

The gross effect of latitude can be evaluated by examining three cases:

A - The flyby path is symmetrical relative to the equator so that the latitude effect in the first half of the flyby, $\theta = 0°$ to $90°$, is exactly offset or balanced by the second half of the flyby, $\theta = 90°$ to $180°$. This case is similar to Figure 1.

B - The flyby path starts at low latitude and finishes at high latitude, Figure 2.

C - The flyby path starts at high latitude and finishes at low latitude, Figure 3.

Per equation *(1)* and Figure 1 in the first half of the flight path the effect of the anomalous, centrally directed acceleration is to increase the speed of the spacecraft whereas the effect in the second half of the flight path is to decrease the spacecraft's speed. By its definition Case A produces no net anomalous acceleration or deceleration of the spacecraft because the first and second halves of the flight path balance and offset each other.

In Case B, the first, i.e. acceleration, half of the flight path is at low latitude where the latitude effect only modestly reduces the anomalous acceleration magnitude. But for that case and path the second, i.e. deceleration, half of the flight path is at a high latitude where the latitude effect greatly reduces the anomalous acceleration magnitude. The net effect is a relatively large acceleration followed by a lesser deceleration for a net increase in the spacecraft's speed.

In Case C, the effect is just the reverse of that in Case B; the first, i.e. acceleration, half of the flight path is at high latitude where the effect of the latitude greatly reduces the anomalous acceleration magnitude. But for that case and path the second, i.e. deceleration, half of the flight path is at a low latitude where the effect of the latitude only modestly reduces the anomalous acceleration magnitude. The net effect is a relatively small acceleration followed by a greater deceleration for a net decrease in the spacecraft's speed.

a. <u>Equatorial View - Flyby</u>

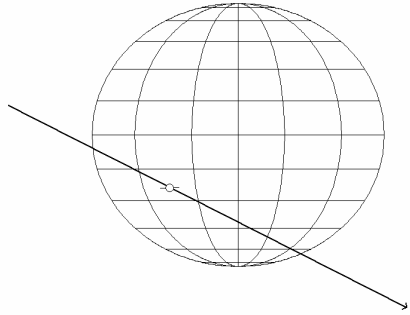

b. <u>Equatorial View - Flyby, Rotated</u>

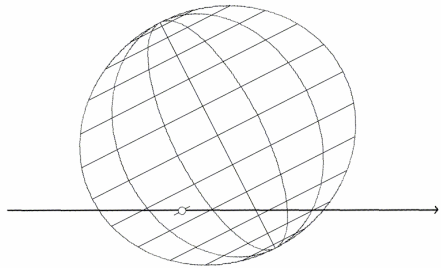

c. <u>Anomalous Acceleration Vectors</u>

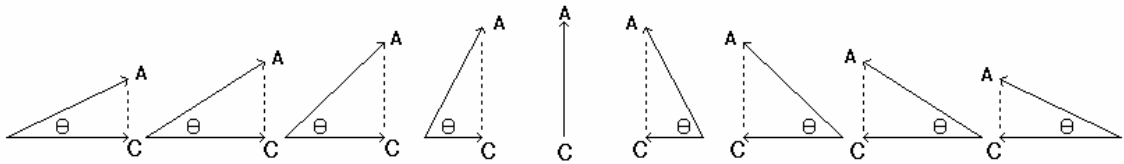

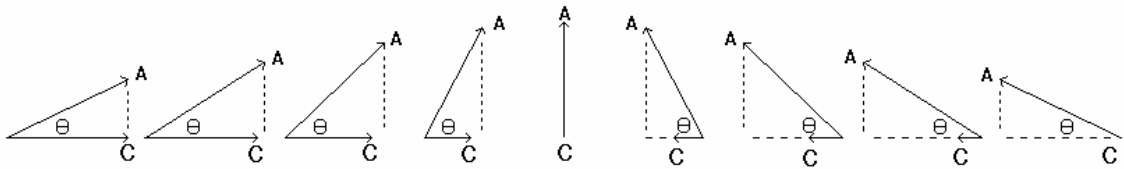

The result in this case is a net acceleration
[to the right in the diagrams].

*Figure 2*
*A Pass at Increasing Latitude*

a. <u>Equatorial View - Flyby</u>

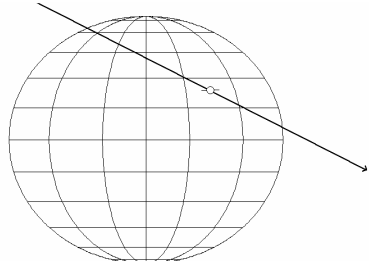

b. <u>Equatorial View - Flyby, Rotated</u>

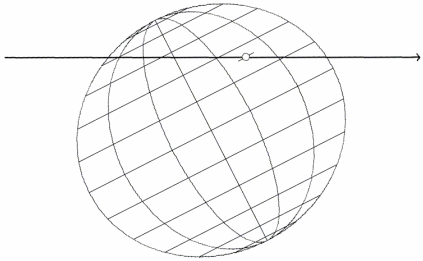

c. <u>Anomalous Acceleration Vectors</u>

Vectors As In zero Latitude Pass Case

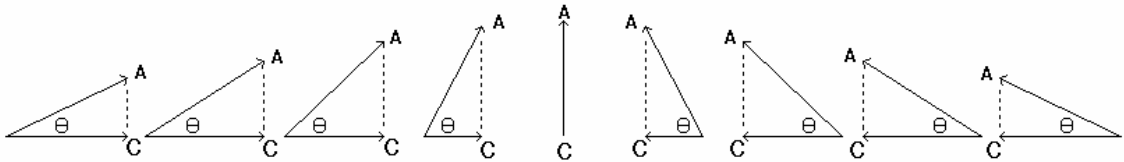

Above Vectors as Further Reduced by Non-Zero Latitude
Reduction Factor = Cosine[Latitude]

| Latitude: | 90° | 80° | 70° | 60° | 50° | 40° | 30° | 20° | 10° |
|---|---|---|---|---|---|---|---|---|---|
| Factor: | .00 | .17 | .34 | .50 | .64 | .77 | .87 | .94 | .99 |

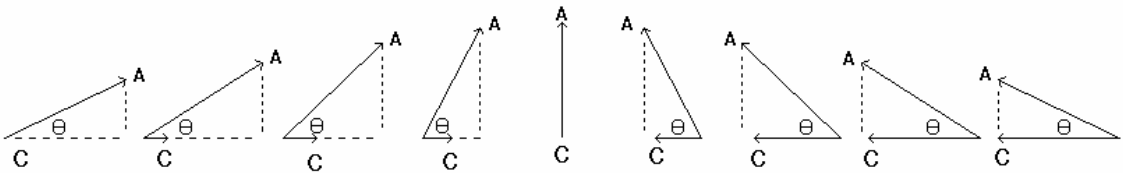

The result in this case is a net deceleration, that is an acceleration toward the left in the diagrams, against the direction of the flyby.

*Figure 3*
*A Pass at Decreasing Latitude*

Therefore, depending on the specific flight path of the spacecraft's flyby pass of Earth the spacecraft may experience an overall net anomalous acceleration or a net anomalous deceleration, those in various amounts depending on the specific encounter, and zero net modification if the path is perfectly symmetrical.

It only remains, then, to determine the cause of centrally directed, distance independent, anomalous accelerations appearing as a near Earth effect [the Flybys Anomaly], a Solar effect [the Pioneer Anomaly] and in general as a cosmic effect.

## *The Cosmic Anomalous Acceleration*

In addition to the relatively recent Pioneer and Flyby anomalies there is another instance of an anomalous centrally directed, distance independent, acceleration [in addition to that of natural gravitation] that has been observed for the better part of a century. That is the indication by galactic rotation curves that galaxies are held together by an unaccounted for, i.e. anomalous, acceleration. That acceleration is centrally directed in each galaxy and is of constant magnitude, independent of distance. It is the same as the anomalous acceleration indicated in the Pioneer Anomaly and it is the same as the anomalous acceleration indicated in the Flybys Anomaly. The existence of those anomalous accelerations is thoroughly and soundly established by large bodies of data and observations.

At the time of the discovery of the anomalous acceleration in galaxies in 1933 and for decades thereafter the occurrence in galaxies was the only occurrence known. The existence of "dark matter" was hypothesized to account for the galaxies' anomalous acceleration by proposing a gravitational cause. To account for its distance independent non-gravitational behavior the distribution of the "dark matter" in a suitably arranged "halo" was proposed.

However, no "dark matter" has ever been detected in spite of substantial efforts. It is only, and remains only, hypothesized to account for the behavior of galactic rotation curves indicating centrally directed acceleration without apparent cause or source.

Thus we have identical anomalous accelerations, small in magnitude, not gravitational, centrally directed, distance independent, occurring: relative to our Earth, relative to our solar system's Sun, acting centrally directed in every rotating galaxy in the cosmos, and even so acting in observed rotating groups of galaxies. What could produce such a phenomenon ? What would cause there to be a universe-wide occurrence of such same accelerations ?

Taken together, planet relative, star relative, galaxy relative, they collectively are a systematic contraction, a gradual reduction in the length component of every physical quantity in the universe.

Objections that such an effect would conflict with the known planetary system performance per the highly accurate planetary ephemeris are a mistaken interpretation of the situation. Consider a planet in circular orbit around a sun as in Figure 4, below.

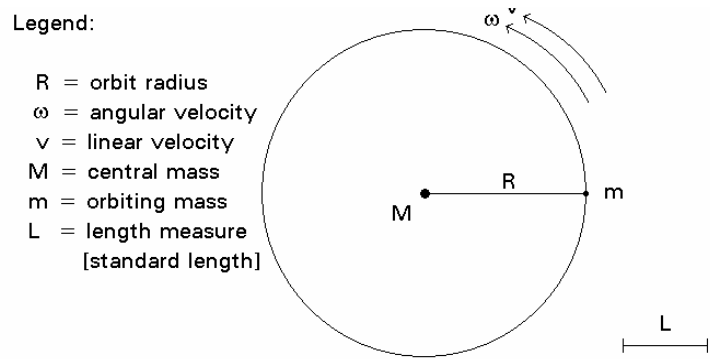

*Figure 4*

The relationship governing the motion is, of course, equation *(4)*, below

*(4)*  Centripetal Acceleration  =  Gravitational Attraction
           Required                    Acceleration

$$V^2/R \text{ (or) } R \cdot \omega^2 = G \cdot M / R^2$$

Now, let the length dimensional aspect [with the dimensions of all quantities expressed in the fundamental dimensions of mechanics, *[L], [M],* and *[T]*] of all quantities decay, becoming gradually smaller with time. That is, let all lengths, *[L]*, decrease by being multiplied by the decay function, *D(t)*, per equation *(3)*, below. [For the present purpose the form of the decay function is irrelevant except that it must be a function of time. The decaying exponential is used because it is common in nature.]

*(5)*  $D(t) \equiv \varepsilon^{-[t/\tau]}$, where $\tau$ is the time constant of the decay

Then the quantities involved in equation *(4)* all change to as follows.

*(6)*  *The Orbital Radius, R, [dimension = **L**]*

$$R \text{ becomes } R(t) = R(t=0) \cdot \varepsilon^{-[t/\tau]}$$

*The Gravitational Constant [dimensions = **L**$^3$/$_{M \cdot T}$ 2]*

$$G \text{ becomes } G(t) = G(t=0) \cdot \left\{ \varepsilon^{-[t/\tau]} \right\}^3$$

*Centripetal Acceleration Required [dimensions = **L**/$_{T}$2]*

$$R \cdot \omega^2 \text{ becomes } R(t) \cdot \omega^2 = \left[ R(t=0) \cdot \varepsilon^{-[t/\tau]} \right] \cdot \omega^2$$
$$= \left[ R(t=0) \cdot \omega^2 \right] \cdot \varepsilon^{-[t/\tau]}$$

or

$$\frac{V^2}{R} \text{ becomes } \frac{[V(t)]^2}{R(t)} = \frac{\left[ V(t=0) \cdot \varepsilon^{-[t/\tau]} \right]^2}{\left[ R(t=0) \cdot \varepsilon^{-[t/\tau]} \right]}$$
$$= \frac{[V(t=0)]^2}{R(t=0)} \cdot \varepsilon^{-[t/\tau]}$$

*Gravitational Attraction Acceleration [dimensions = **L**/$_{T}$2]*
    [and where the G dimensions = **L**$^3$/$_{M \cdot T}$ 2]

$$\frac{G \cdot M}{R^2} \text{ becomes } \frac{G(t) \cdot M}{[R(t)]^2} = \frac{\left[ G(t=0) \cdot \left\{ \varepsilon^{-[t/\tau]} \right\}^3 \right] \cdot M}{\left[ R(t=0) \cdot \varepsilon^{-[t/\tau]} \right]^2}$$
$$= \frac{G(t=0) \cdot M}{[R(t=0)]^2} \cdot \varepsilon^{-[t/\tau]}$$

The overall net effect is: *R* decreases, the required centripetal acceleration decreases in proportion, the gravitational attraction likewise decreases in proportion, and $\omega$ is unchanged.

Furthermore, we observers, using our measuring standard ruler, length *L* of the above Figure 2, would never detect any of the decay because our standard length would also be decaying at exactly the same rate, in the same proportion.

The point of this obvious mathematics / physics exercise is that a universal decay of the length aspect of all material reality would not conflict with the planetary ephemeris and would not even be detectable at all except in unusual circumstances such as the Pioneer and Flyby anomalies and the evidence of galactic rotation curves.

Returning to the orbiting body of Figure 4, reproduced as Figure 5 below, the figure's annotations slightly modified, the development of the anomalous acceleration is very direct.

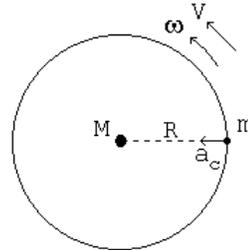

```
Legend:
    R = orbit radius
    ω = angular velocity
    V = linear velocity
    M = central mass
    m = orbiting mass
    a_c = centripetal acceleration
        = Newtonian gravitation + anomalous acceleration a_p
```

*Figure 5*

The Newtonian component of the centripetal acceleration is only sufficient to maintain the orbit, to keep $R$ constant, to prevent its increasing. For the orbiting body, $m$, to gradually approach the central mass, $M$, that is for $R$ to decrease, additional inward acceleration is required. That acceleration is the anomalous acceleration. It is an unavoidable concomitant effect of the universal decay or contraction of the length dimension $[L]$ of $R$ in the above example and of all material reality.

Details on the universal decay -- its cause, origin and characteristics are too lengthy for this report and are provided in full in references [2] and [3].

Experiments to test the universal decay and measure its parameters are proposed in detail in reference [2] and briefly in reference [3].